\documentclass[aps,preprint]{revtex4}%
\usepackage{amsfonts}
\usepackage{amsmath}
\usepackage{amssymb}
\usepackage{graphicx}%
\begin{document}
\preprint{HEP/123-qed}
\title[Effect of the break-up on ...]{Effect of the break-up on the fusion and elastic scattering of weakly bound
projectiles on $^{64}$Zn}
\author{P. R. S. Gomes$^{1}$, M.D. Rodr\'{\i}guez$^{2}$, G.V. Mart\'{\i}$^{2}$, , I.
Padron$^{1}$\thanks{Permanent address: Center of Applied Studies to Nuclear
Development, Havana, Cuba, P.O. Box 6122}, L.C. Chamon$^{3}$, J.O.
Fern\'{a}ndez Niello$^{2,4}$, O.A. Capurro$^{2}$, A.J. Pacheco$^{2}$, J.E.
Testoni$^{2}$, A. Arazi$^{2}$, M. Ram\'{\i}rez$^{2}$, R. M. Anjos$^{1}$, J.
Lubian$^{1}$, R. Veiga$^{1}$, R. Liguori Neto$^{3}$, E. Crema$^{3}$, N.
Added$^{3}$, C. Tenreiro$^{5}$and M.S. Hussein$^{6}$.}
\affiliation{$^{1}$Instituto de F\'{\i}sica, Universidade Federal Fluminense, Av. Litoranea
s/n, Gragoat\'{a}, Niter\'{o}i, R.J.,24210-340, Brazil. $^{2}$ Laboratorio
Tandar, Departamento de F\'{\i}sica, Comisi\'{o}n Nacional de Energ\'{\i}a
At\'{o}mica, Av. del Libertador 8250, (1429), Buenos Aires, Argentina. $^{3}%
$Departamento de F\'{\i}sica Nuclear, Universidade de S\~{a}o Paulo, Caixa
Postal 66318, 05315-970, S\~{a}o Paulo, S.P., Brazil. $^{4}$Escuela de Ciencia
e Tecnolog\'{\i}a, Universidad de San Mart\'{\i}n, Argentina. $^{5}$Facultad
de Ingenier\'{\i}a, Universidad de Talca,Curic\'{o}, Chile. $^{6}$Departamento
de F\'{\i}sica Matem\'{a}tica, Universidade de S\~{a}o Paulo, Caixa Postal
66318, 05315-970, S\~{a}o Paulo, S.P., Brazil}

\begin{abstract}
We study the behavior of the fusion, break-up, reaction and elastic scattering
of different projectiles on $^{64}$Zn, at near and above barrier energies. We
present fusion and elastic scattering data with the tightly bound $^{16}$O and
the stable weakly bound $^{6}$Li, $^{7}$Li and $^{9}$Be projectiles. The data
were analyzed by coupled channel calculations. The total fusion cross sections
for these systems are not affected by the break-up process at energies above
the barrier. The elastic (non-capture) break-up cross section is important at
energies close and above the Coulomb barrier and increases the reaction cross
sections. In addition we also show that the break-up process at near and
sub-barrier energies is responsible for the vanishing of the usual threshold
anomaly of the optical potential and give rise to a new type of anomaly.

\end{abstract}
\volumeyear{year}
\volumenumber{number}
\issuenumber{number}
\eid{identifier}
\date[Date text]{date}
\received[Received text]{date}

\revised[Revised text]{date}

\accepted[Accepted text]{date}

\published[Published text]{date}

\startpage{101}
\endpage{102}
\maketitle

\section{Introduction}

The fusion of weakly bound nuclei, both stable and radioactive, has been the
subject of intense theoretical and experimental
activities\cite{1,2,3,4,5,6,7,8,9,10,11,12,13,14,15,16,17,18,19,20,21,22,23,24,25,26,27,28,29,30,31,32,33,34,35,36,37,38,39,40,41,42,43,44,45,46,47,48}%
. Since the beginning of the 1990's, theorists have been facing conflicting
ideas about whether the fusion of such nuclei, of great importance in nuclear
astrophysics, is enhanced or hindered at low energies owing to the strong
coupling to the break-up channel\cite{10,11,12,13,14,15,16,17,18,19,20,21,22}.
The reason for such divergent conclusions has been the lack of a trust-worthy
continuum-coupled channels theory. Even recently, when the
Continuum-Discretized Coupled Channel (CDCC) theory has been utilized for the
calculation of fusion\cite{21,22}, the results are not fully satisfactory due
to the difficulty in calculating the incomplete fusion (the fusion of a piece
of the projectile) contribution to the total fusion\cite{21}.

One of the motivations in studying the fusion of stable weakly bound nuclei
with heavy targets is the considerably high intensity of such beams as
compared to radioactive beams. Most of the essential features of the
phenomenon, namely the importance of the break-up channel even at low
energies, the related significance of incomplete fusion, and the fact that
what is normally measured is the total fusion (the sum of complete and
incomplete fusion), are present in both cases, albeit less conspicuously in
the stable isotope case. The discerning of the sought after complete fusion
cross section constitutes the real experimental challenge in the field. The
recent publication on the $^{6}$He + $^{238}$U system\cite{23} is an important
step in this direction. Yet, detailed investigation of the fusion of stable
weakly bound nuclei is still very important as it allows the study of a wider
range of systems and the development of much needed systematics.

The suitable stable nuclei for this kind of study are $^{9}$Be, $^{6}$Li and
$^{7}$Li which have threshold break-up energies between 1.48 MeV and 2.45 MeV.
The full understanding of the fusion and break-up processes induced by these
beams is an important reference for similar studies involving radioactive
proton and neutron rich projectiles\cite{7,8,9}.

Among the questions on this subject that have to be answered, one finds: What
are the values of $\sigma_{EBU}$ (EBU or elastic break-up is the break-up not
followed by the capture of any of the break-up fragments by the target) for
different energy regimes, target masses and threshold break-up energies? Does
the break-up affect the fusion cross section or just increase the reaction
cross section? If it affects the fusion process, does it enhance or suppress
the fusion cross section at different energy regimes and target masses? Is the
effect on the complete fusion or total fusion, the last one corresponding to
the sum of the normal two-body fusion with the fusion processes following
break-up (ICF (incomplete fusion) and CFBU (complete fusion following break-up))?

\qquad There are several theoretical aspects to be considered when one wants
to study the influence of the break-up process on the fusion cross section.
One should include all the mentioned different reaction mechanisms related
with the break-up and the relative motion of the fragments and their
interactions, the boundary conditions used for the occurrence of complete
fusion, such as the distance where the fusion is decided, and definitions of
CF and ICF related to bound and unbound states. Calculations have to be
performed by considering bound states-continuum couplings, with or without
resonance states, discrete continuum states and continuum-continuum couplings,
Coulomb and nuclear excitations have to be considered, as well as their
coherent interferences. Up to now, there is no such a complete theoretical
study in this field. Different approaches have been used, such as coupled
channel calculations (CCC) that do not take into account the break-up
process\cite{19}, continuum discrete coupled channel calculations
(CDCC)\cite{20,21,22}, semi-classical trajectories, survival probability
concept and dynamic polarization potentials\cite{10,11,12,13,14,15,16,17,18}%
.\qquad

\qquad Experimentally, it is important to know if it is possible to separate
the CF and ICF processes. Usually the residues following both processes are
very similar or identical, and therefore the measurement of residues using
different techniques and/or particle identification devices (i.e., charged
particle detectors, time of flight detectors, ionization chamber, x-rays
delayed, etc.) is not able to distinguish between them. This is even more
dramatic for light systems for which the main evaporation channels include
charged particles (protons and alphas). Therefore, most of the available data
in the literature correspond to total fusion (TF) cross section, although
there are reports of some measurements of CF and ICF
separately\cite{24,25,26,27,28,29,30}. The measurement of EBU cross sections
requires difficult exclusive experiments\cite{31,32,33,34,35,36}.

\qquad The present paper is organized as follows. In Section 2, we present the
experimental details on the low energy induced reactions of $^{16}$O (stable,
tightly bound), $^{6}$Li, $^{7}$Li and $^{9}$Be (stable, weakly bound)
projectiles on $^{64}$Zn. In Section 3 we give the experimental results for
the fusion and the elastic scattering angular distributions, through which the
reaction cross sections are extracted and discussed. In Section 4, the elastic
scattering angular distributions are analyzed with the S\~{a}o Paulo Optical
Potential (SPOP), which takes into account the effect of the nonlocality
within the double folding prescription. A new threshold anomaly directly
related to the persistent action of the coupling to the break-up channel, even
at below barrier energies, is also discussed in this section. Finally, in
Section 5, we present our concluding remarks. A short version of a part of
this work has already been published\cite{37}.

\section{Experimental details}

\qquad The experiments to measure the elastic scattering for the $^{16}%
$O+$^{64}$Zn, $^{9}$Be + $^{64}$Zn and $^{6,7}$Li+$^{64}$Zn systems were
performed at the Pelletron Laboratory of the Universidade de S\~{a}o Paulo
(USP), S\~{a}o Paulo, Brazil. The elastic scattering cross sections were
measured with a set of nine surface barrier detectors, placed at 40 cm from
the target, with 5$^{0}$ angular separations between two adjacent detectors
having a resolution of the order of 350 keV at 20 MeV. In front of each
detector, there was a set of circular collimators and slits for the definition
of the solid angles and to avoid slit-scattered particles. The experimental
array has been previously described in detail in references\cite{38,39,40}.
The angle determination was made by reading on a goniometer with a precision
of 0.5$^{0}$. An additional surface barrier detector used as monitor was
placed at 20$^{0}$, relative to the beam direction, for normalization
purposes. For the Li isotopes scattering, the relative solid angles of the
detectors were determined by Rutherford scattering of $^{6}$Li on the $^{197}%
$Au thin backing layer deposited onto the target. The $^{64}$Zn target had a
thickness of 60 $\mu$g/cm$^{2}$. Three beam energies were used for the $^{6}%
$Li beam, 17, 20 and 22 MeV, and two energies for the $^{7}$Li beam, 20 and 22
MeV, with the angular range 10$^{0}$ $\leq$ $\theta_{Lab}$%
\protect\rule{0.1in}{0.1in}
$\leq$ 150$^{0}$. The uncertainties in the differential cross section data
vary from 1$\%$ to 8$\%$. The measurements with the $^{9}$Be and $^{16}$O beam
have been previoulsly reported\cite{38,39}.

\qquad The total fusion cross sections of the $^{6}$Li and $^{9}$Be + $^{64}%
$Zn systems were measured using the experimental facilities of the TANDAR
Laboratory, in Buenos Aires, Argentina. Beams of $^{6}$Li at 16, 18, 20, 22
and 24 MeV, and $^{9}$Be at 20, 22 and 24 MeV were delivered by the 20
UD-tandem accelerator. A metallic $^{64}$Zn target, with thickness of 50 $\mu
$g/cm$^{2}$, was deposited on a 10 $\mu$g/cm$^{2}$ carbon backing and used for
runs collecting data for both measured systems. The fusion cross sections data
were obtained using the time of flight method and complement previously
reported results obtained by the same method, at higher energies, for the
$^{6}$Li + $^{64}$Zn system\cite{41} and by the gamma ray spectroscopy method
for the $^{9}$Be + $^{64}$Zn system\cite{38}. Further description of the
detection system and additional experimental details can be found in Ref. 42.
Figure 1 shows a typical bi-parametric energy vs. mass spectrum and the
projection of this spectrum on the mass axis, for the $^{9}$Be + $^{64}$Zn
system, measured at $\theta_{Lab}$ = 10$^{0}$ and 24 MeV beam energy.

\section{Experimental results for fusion and reaction cross sections and
discussions}

\qquad Table 1a) shows the results for the fusion and the deduced reaction
cross sections obtained for the $^{6}$Li, $^{7}$Li + $^{64}$Zn systems,
together with previously reported fusion data\cite{41}. In Table 1b) we quoted
the results of measured fusion and derived reaction cross sections for the
$^{9}$Be + $^{64}$Zn system from the present work and from Ref. 38, derived
reaction cross sections from Ref. 39 for the $^{16}$O + $^{64}$Zn system and
the previously reported fusion cross section for the $^{16}$O + $^{64}$Zn
system\cite{43}. It is worthwhile to notice that the total fusion cross
sections obtained for the $^{9}$Be + $^{64}$Zn system by the gamma ray
method\cite{38,44} and using the TOF technique agree very well, as can be seen
in Figure 2.

In order to study the possible influence of the break-up of weakly bound
nuclei on the fusion and reaction cross sections, we use $^{16}$O + $^{64}$Zn
as a reference system since it has a negligible break-up cross section. Figure
3 shows that total fusion ($\sigma_{TF}$) and reaction cross sections
($\sigma_{R}$) are similar for the whole energy range. The dashed line is the
result of CCFULL \cite{19} calculations without any couplings, and the full
line is the result including the couplings of the first two excited states of
the target (2$_{1}^{+}$ and 2$_{2}^{+}$ states at E*(2$_{1}^{+}$)=0.792 MeV
and E*(2$_{2}^{+}$)=1.799 MeV, respectively). Almost no difference between the
two calculations can be observed, as expected at this energies above the
Coulomb barrier. A good fit of the total fusion excitation function is obtained.

Figure 4 shows that for the weakly bound projectile $^{9}$Be, $\sigma_{TF}$
and the deduced $\sigma_{R}$ are also very similar for most of the studied
energy range, but for energies close to the Coulomb barrier, $\sigma_{TF}$
becomes appreciably smaller than $\sigma_{R}$. Here, $\sigma_{TF}$ corresponds
to the sum of the fusion of $^{9}$Be, $^{8}$Be and one alpha fragment with the
target. Since the inelastic excitation of $^{64}$Zn has small cross
section\cite{38}, and supposing that transfer reaction cross sections are also
negligible, we conclude from the unitary constrainty $\sigma_{R}$ $\geq$
$\sigma_{TF}$ $+$ $\sigma_{EBU}$ that the cross section of the $^{9}$Be
break-up into two alpha particles plus one neutron without any capture (EBU)
is significative, when compared with $\sigma_{TF}$, only at energies close to
the barrier. The full and dashed lines in Figure 4 correspond to CCFULL
calculations for the total fusion, performed at similar way as in Figure 3.
The derived $\sigma_{EBU}$ values are also shown in Figure 4. Small error bars
are obtained for the two lowest energies, since $\sigma_{EBU}$ was derived by
the difference between the large $\sigma_{R}$ value and the relatively small
$\sigma_{TF}$ value estimated from the CCFULL calculations, whereas for higher
energies, the error bars are very large due to the uncertainties coming out
from the difference between two large numbers, since $\sigma_{R}$ $\sim$
$\sigma_{TF}$ at this energy regime. A good fit of the total fusion excitation
function is obtained, showing that there is no total fusion suppression or
enhancement, compared with predictions from the bare potential of the CCFULL
code. Therefore, we conclude that the break-up of $^{9}$Be does not affect
$\sigma_{TF}$ , but rather increases $\sigma_{R}$ at energies close to the
barrier. From the fusion data obtained by the gamma ray spectroscopy method it
was possible to estimate\cite{37,38} that the $\sigma_{ICF}$ is less 10$\%$ of
the $\sigma_{TF}$ , where $\sigma_{ICF}$ corresponds to the fusion of one
alpha particle fragment with the target\cite{38,44}.

Figure 5 shows $\sigma_{TF}$ and the derived $\sigma_{R}$ for the $^{6}$Li +
$^{64}$Zn system. The total fusion excitation function is well described by
CCFULL calculations, without couplings or including the coupling to the first
two excited states of the target, leading to the conclusion that $\sigma_{TF}$
is not affected by the presence of the break-up process. It is worth mention
that our $\sigma_{TF}$ data are very similar to those for the $^{6}$Li +
$^{59}$Co system\cite{45}, and the later were well described by CDCC
calculations\cite{46}. Contrary to the behavior with the $^{9}$Be projectile,
$\sigma_{R}$ for the $^{6}$Li are larger than $\sigma_{TF}$ even at the
highest energies, and therefore $\sigma_{EBU}$ has significant cross section
at this regime.

Figure 6 shows $\sigma_{TF}$ and $\sigma_{R}$ for the $^{7}$Li + $^{64}$Zn
system. As there are no $\sigma_{TF}$ data available at low energies, we did
not attempt to perform coupled channel calculations. The full line is the
result of CCFULL calculations without any coupling. Although the two energies
for which $\sigma_{R}$ were derived are slightly lower than the ones for which
$\sigma_{TF}$ are available, one can notice that $\sigma_{R}$ $>$ $\sigma
_{TF}$ even at energies above the barrier, similar to the behavior with the
$^{6}$Li projectile. The derived $\sigma_{EBU}$ for this system are shown in
Figure 6, where the $\sigma_{TF}$ value used was obtained from the CCFULL predictions.

From the behavior of the derived $\sigma_{TF}$ , $\sigma_{R}$ and
$\sigma_{EBU}$ for the $^{9}$Be, $^{6}$Li, $^{7}$Li + $^{64}$Zn systems and
also using the fact that for the $^{6}$He + $^{64}$Zn system, $\sigma_{R}$
$>>$ $\sigma_{TF}$ at energies above the barrier\cite{37,47,48}, we draw a
scenario as follows: At energies above the barrier, the weakly bound nuclei
$^{6}$He and $^{6,7}$Li break-up at relatively large distances from the
target, and their fragments move in different directions, leading to
significant $\sigma_{EBU}$ (or, alternatively, the EBU corresponding to large
partial waves has significant cross section). In a different break-up process,
the $^{9}$Be breaks-up into $^{8}$Be and one neutron, the $^{8}$Be moves
almost in the same direction as the $^{9}$Be, and only sometime later ($\sim
$10$^{-16}$ sec) the two alpha particles are produced. The whole process for
the $^{9}$Be break-up, particularly with a relatively light target with a not
very strong Coulomb potential, reduces the probability of EBU at this regime.
At energies close and below the barrier, one expects that $\sigma_{EBU}$
becomes significant for any of the weakly bound projectiles, particularly if
the target is heavy and the Coulomb break-up takes place at relatively larger distances.

\section{The elastic scattering analyzed with the S\~{a}o Paulo potential}

The strong presence of the break-up process at near barrier energies affects
the elastic scattering and reaction cross section of weakly bound nuclei in
such way that the usual threshold anomaly of the optical potential is no
longer present, except from $^{7}$Li induced reactions, as it was previoulsly
reported\cite{30,40,49,50,51,52,53,54}. For the $^{9}$Be + $^{64}$Zn system,
in previous works\cite{38,44}, we followed the usual procedure of obtaining
the energy-dependence of the real and imaginary parts of the optical potential
at the strong absorption radius, using a Woods-Saxon shape for the real and
volume imaginary potentials and a derivative Woods-Saxon shape for the surface
imaginary potential. The presence of the break-up channel at these low
energies does not allow the vanishing of the imaginary part of the potential
as the energy approaches the barrier. Indeed, an increase in the imaginary
potential is observed at the lowest energy, as it was also observed by
Signorini et al. for the $^{9}$Be + $^{209}$Bi system\cite{53}. On the other
hand, for the tightly bound $^{16}$O + $^{64}$Zn system the usual threshold
anomaly is present: the imaginary potential decreases when the bombarding
energy decreases towards the barrier\cite{39}.

However, the threshold anomaly was recently observed for the $^{9}$Be +
$^{208}$Pb system\cite{54}, showing that the behavior of the elastic
scattering of weakly bound nuclei is not yet fully understood. Therefore, we
decided to study the threshold anomaly by an alternative approach, that uses a
global parameter-free optical potential known as S\~{a}o Paulo potential
(SPOP) \cite{55,56}. Very recently, we have applied the same procedure for the
study of the elastic scattering of the $^{9}$Be + $^{27}$Al system\cite{52}.
This potential is based on the Pauli nonlocality involving the exchange of
nucleons between projectile and target. Within this model, the nuclear
interaction is connected with the folding potential V$_{F}$
through\cite{55,56} the following formula:%

\begin{equation}
V_{N}(R,E)\approx V_{F}(R)exp(-4v^{2}/c^{2}),
\end{equation}
where $c$ is the speed of light in vacuum and $v$ is the local relative
velocity between the two nuclei. In this context, and considering an extensive
systematics of nuclear densities, a systematization of the imaginary part of
the optical potential was also obtained:%

\begin{equation}
W(R,E)=N_{i}V_{N}(R,E).
\end{equation}

\qquad For several systems, elastic scattering angular distributions over wide
energy ranges were simultaneously well fitted with an optical potential
defined by Eqs. (1) and (2), with N$_{i}$ = 0.78 \cite{56}.

In order to explain the elastic scattering angular distributions for the
$^{9}$Be + $^{64}$Zn system, we started with the SPOP without any free
parameter (and consequently no data fit), corresponding to the use of Equation
(1), multiplied by N$_{R}$ = 1.0, and Equation (2) with N$_{i}$ = 0.78
\cite{56}. The results are shown in Figure 7 by dashed lines. Then, the values
of N$_{R}$ and N$_{i}$ were considered as free parameters to fit the data. The
solid lines in Figure 7 correspond to the best data fits. The results of the
energy dependence of the best N$_{R}$ and N$_{i}$ values are shown in Figure
8. The derivation of the error bars was done as follows: we defined the
maximum acceptable $\chi^{2}$ value as $\chi_{Max}^{2}$ = $\chi_{Min}^{2}$ +
$\chi_{Min}^{2}$ /N, where N is the number of points of the angular
distribution; and then we found the range of N$_{R}$ and N$_{i}$ corresponding
to $\chi^{2}$ smaller or equal to $\chi_{Max}^{2}$.

From Figure 8 one can observe a dramatic deviation of the energy dependence
from the so-called threshold anomaly (TA). The value of N$_{R}$ is roughly 0.9
down to the barrier, where one sees a significant drop at E
$<$
V$_{B}$. Thus the coupling to the break-up channel results in an overall
repulsion. The behavior of N$_{i}$, also shown in Figure 8, is even more
dramatic: the value 0.78 seems reasonable up to energies in the vicinity of
the barrier. As the energy is lowered further, a significant increase in
N$_{i}$ is observed. This is in sharp contrast to the threshold anomaly
behavior seen in the scattering of tightly bound nuclei where one finds the
exact opposite behavior of N$_{i}$ and N$_{R}$ at E
$<$
V$_{B}$. The underlying theoretical tool behind the TA is the dispersion
relation that relates the real part of the dynamic polarization potential
(DPP) - the Feshbach potential - to an energy integral involving the imaginary
part. We did not calculate this potential here, but by fitting the data with a
slightly modified SPOP, we are effectively taking into account the DPP.

What we are observing here is a new type of threshold anomaly directly linked
to the coupling to the break-up channel. We call this the Break-up Threshold
Anomaly (BTA). The repulsive nature of the real part of the break-up DPP has
been discussed by several authors\cite{57,58,59,60}. The fact that the
threshold for break-up extends far down to low energies in the case of weakly
bound nuclei is clearly manifested in the BTA. Being repulsive in nature, the
break-up DPP would lead to an increase in the barrier height resulting in a
decrease of the complete fusion cross section at sub-barrier energies. This
should become more conspicuous in loosely bound unstable nuclei. Of course the
total fusion, being the sum of incomplete fusion and complete fusion, would
not be affected by the DPP, which was borne out clearly in the present paper.
A full discussion of this new phenomenon, the BTA, will be reported elsewhere.

For the purpose of comparison, we have followed a similar procedure for the
tightly bound nucleus $^{16}$O scattering from $^{64}$Zn. Figure 9 shows the
best data fits (full lines) and the predictions from the SPOP with N$_{R}$=1
and N$_{i}$=0.78 (dashed lines). The results of the energy dependence of the
best N$_{R}$ and N$_{i}$ values are shown in Figure 10. Here, the usual TA is
observed: as the energy is lowered below the barrier (the threshold of
non-elastic processes), N$_{i}$ , which is about 0.6 at energies above the
barrier, suffers a significant decrease, accompained by a rather sharp
increase in N$_{R}$ (more attraction). This is fully consistent with the
prediction of the dispersion relation.

\section{Summary and conclusions}

\qquad Total fusion cross sections at 16, 18, 20, 22 and 24 MeV projectile
energies were measured for the $^{6}$Li+$^{64}$Zn system and at 20, 22 and 24
MeV for the $^{9}$Be+$^{64}$Zn system. In addition, reaction cross sections
were also derived for some energies using the present data and previously
measured data which include those obtained at energies above the barrier for
the reaction $^{7}$Li+$^{64}$Zn. For the $^{16}$O+$^{64}$Zn system, we were
able to derive reaction cross sections in a wide energy range (40-68.5 MeV)
using elastic scattering data previously reported.

\qquad We conclude that for the fusion of $^{9}$Be with the medium mass target
$^{64}$Zn, the TF and the sum of the contributions of $^{9}$Be and $^{8}$Be to
fusion are not affected by the break-up process, at least within the
experimental uncertainties. The $\alpha$-ICF cross section seems to be very
small and therefore can be ignored. The EBU contribution is important only at
near and sub-barrier energies. In the reactions involving projectiles $^{6,7}%
$Li and $^{6}$He, with the same target, the $\sigma_{R}$ and the $\sigma
_{EBU}$ are large at all energies, and both increase as the break-up threshold
energy gets smaller. However, the total fusion cross-section is not affected
by the break-up channel.

The energy dependence of the real and imaginary parts of the potential was
studied by the double folding global S\~{a}o Paulo potential, and the results
indicate the operation of a new type of threshold anomaly in the elastic
scattering of the weakly bound nucleus $^{9}$Be: The imaginary potential
suffers a significant rise as the energy is lowered below the barrier,
accompanied by a sharp decrease of the real potential. This favors an increase
in the barrier height and should result in a decrease in sub-barrier complete
fusion. This new, Break-up Threshold Anomaly (BTA), is in sharp contrast to
the usual threshold anomaly (TA) seen in the energy dependence of the
potential in the case of the elastic scattering of the tightly bound $^{16}$O.
The existence of the BTA is believed to be due to the presence of the strong
coupling to the break-up channel, with large cross section at energies close
and below the Coulomb barrier, and therefore the imaginary potential does not
decrease, in fact it increases, as the energy decreases towards the Coulomb barrier.

\textbf{Acknowledgements}

The authors would like to thank the CNPq, CAPES, FAPERJ, FAPESP and CONICET
for their financial support.

\newpage

\bigskip

Table 1) Fusion and reaction cross sections for the different systems measured
by our groups. $\sigma_{R}$ were obtained from elastic scattering data.

(a) $^{6}$Li and $^{7}$Li + $^{64}$Zn%

\[%
\begin{tabular}
[c]{|c|c|c|c|c|}\hline
E$_{Lab}$ (MeV) & $\sigma_{fus}$ ($^{6}$Li) (mb) & $\sigma_{R}$ ($^{6}$Li)
(mb) & $\sigma_{fus}$ ($^{7}$Li) (mb) & $\sigma_{R}$ ($^{7}$Li) (mb)\\\hline
16.0 & 114 $\pm$ 13 &  &  & \\\hline
17.0 &  & 533 &  & \\\hline
18.0 & 145 $\pm$ 14 &  &  & \\\hline
20.0 & 332 $\pm$ 25 & 878 &  & 854\\\hline
22.0 & 488 $\pm$ 41 & 1094 &  & 1100\\\hline
24.0 & 634 $\pm$ 48 &  & 656 $\pm$ 56 \cite{41} & \\\hline
28.0 & 823 $\pm$ 59 \cite{41} &  & 883 $\pm$ 66 \cite{41} & \\\hline
31.0 & 869 $\pm$ 60 \cite{41} &  & 922 $\pm$ 64 \cite{41} & \\\hline
34.0 & 984 $\pm$ 68 \cite{41} &  & 1002 $\pm$ 69 \cite{41} & \\\hline
37.0 & 1053 $\pm$ 71 \cite{41} &  & 1134 $\pm$ 77 \cite{41} & \\\hline
40.0 & 1022 $\pm$ 65 \cite{41} &  & 1105 $\pm$ 75 \cite{41} & \\\hline
43.0 & 1166 $\pm$ 71 \cite{41} &  & 1254 $\pm$ 81 \cite{41} & \\\hline
\end{tabular}
\]

\bigskip

(b) $^{9}$Be and $^{16}$O + $^{64}$Zn%

\[%
\begin{tabular}
[c]{|c|c|c|c|c|}\hline
E$_{Lab}$ (MeV) & $\sigma_{fus}$ ($^{9}$Be) (mb) & $\sigma_{R}$ ($^{9}$Be)
(mb) & $\sigma_{fus}$ ($^{16}$O) (mb) & $\sigma_{R}$ ($^{16}$O) (mb)\\\hline
17.0 &  & 68 \cite{38} &  & \\\hline
19.0 &  & 199 \cite{38} &  & \\\hline
20.0 & 140 $\pm$ 18 &  &  & \\\hline
21.0 & 358 $\pm$ 35 \cite{38} & 424 \cite{38} &  & \\\hline
22.0 & 472 $\pm$ 46 &  &  & \\\hline
23.0 & 570 $\pm$ 57 \cite{38} & 590 \cite{38} &  & \\\hline
24.0 & 747 $\pm$ 92 &  &  & \\\hline
26.0 & 930 $\pm$ 92 \cite{38} & 871 \cite{38} &  & \\\hline
28.0 &  & 1013 \cite{38} &  & \\\hline
29.0 & 1120 $\pm$ 112 \cite{38} &  &  & \\\hline
40.0 &  &  & 60 $\pm$ 8 \cite{43} & 59.3 (39*)\\\hline
41.0 &  &  &  & 78.9 (39*)\\\hline
42.5 &  &  & 164 $\pm$ 17 \cite{43} & 191 (39*)\\\hline
43.5 &  &  &  & 231 (39*)\\\hline
44.0 &  &  &  & 262 (39*)\\\hline
48.0 &  &  &  & 542 (39*)\\\hline
50.0 &  &  & 536 $\pm$ 60 \cite{43} & 640 (39*)\\\hline
52.0 &  &  &  & 727 (39*)\\\hline
54.0 &  &  &  & 828 (39*)\\\hline
56.0 &  &  &  & 914 (39*)\\\hline
60.0 &  &  & 1095 $\pm$ 110 \cite{43} & \\\hline
62.0 &  &  &  & 1120 (39*)\\\hline
64.0 &  &  &  & 1182 (39*)\\\hline
68.5 &  &  & 1354 $\pm$ 162\cite{43} & 1350 (39*)\\\hline
\end{tabular}
\
\]

(39*)- Reaction cross sections derived from elastic scattering data reported
in Ref 39.

\newpage%

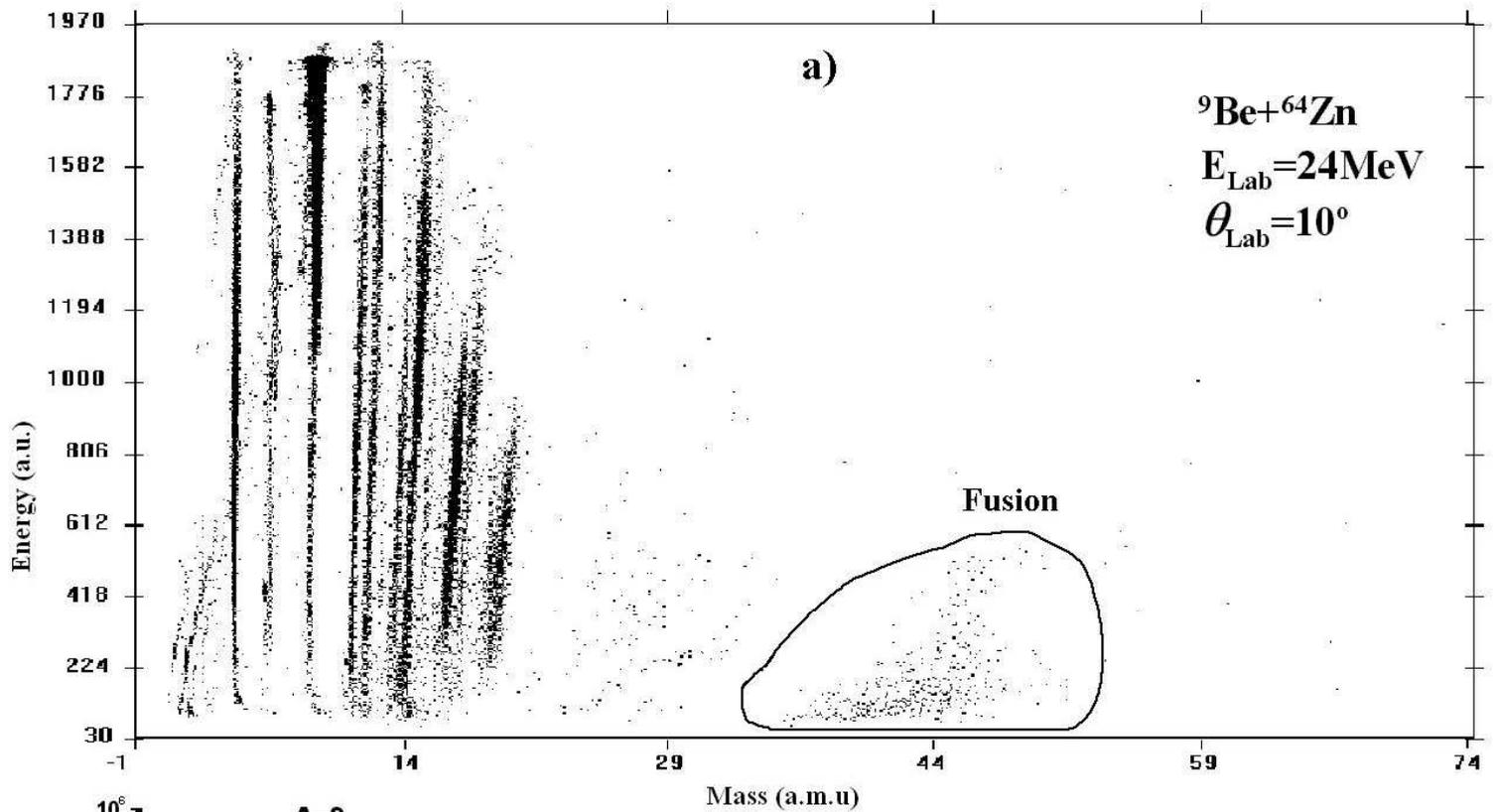
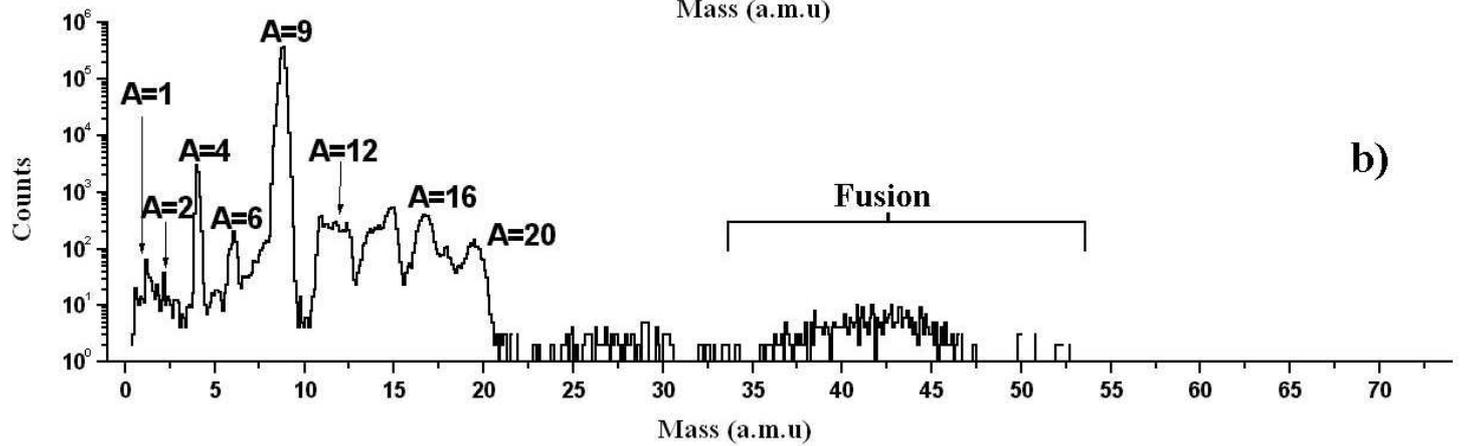
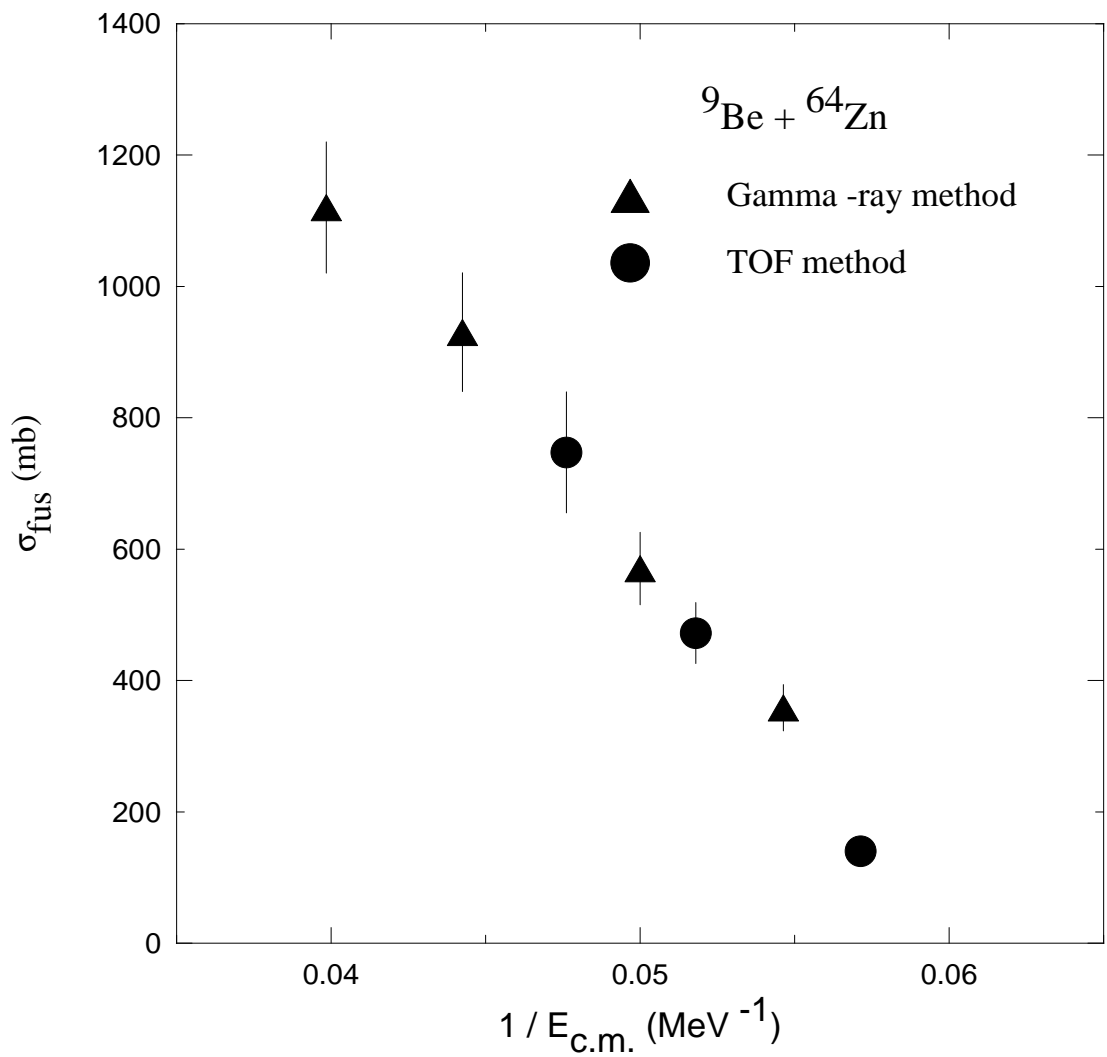
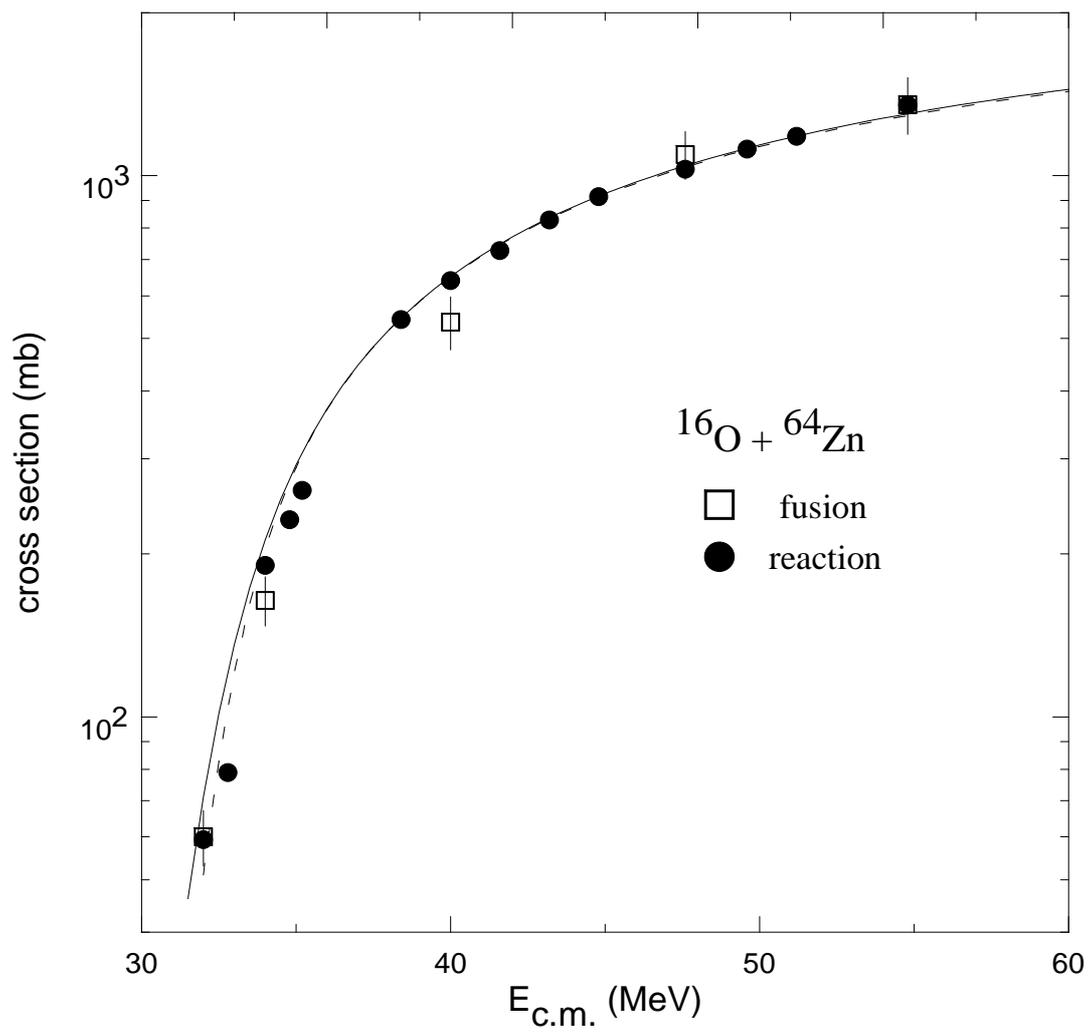
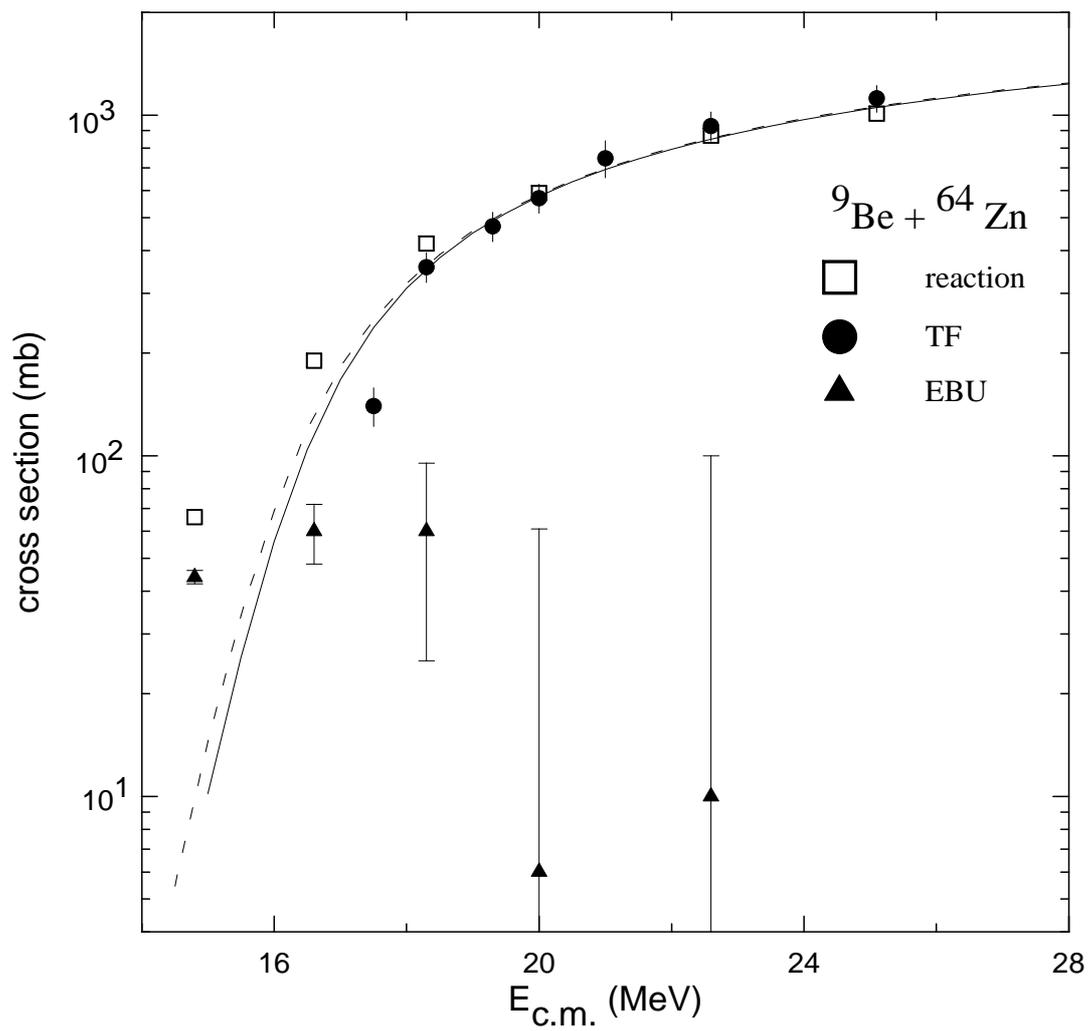
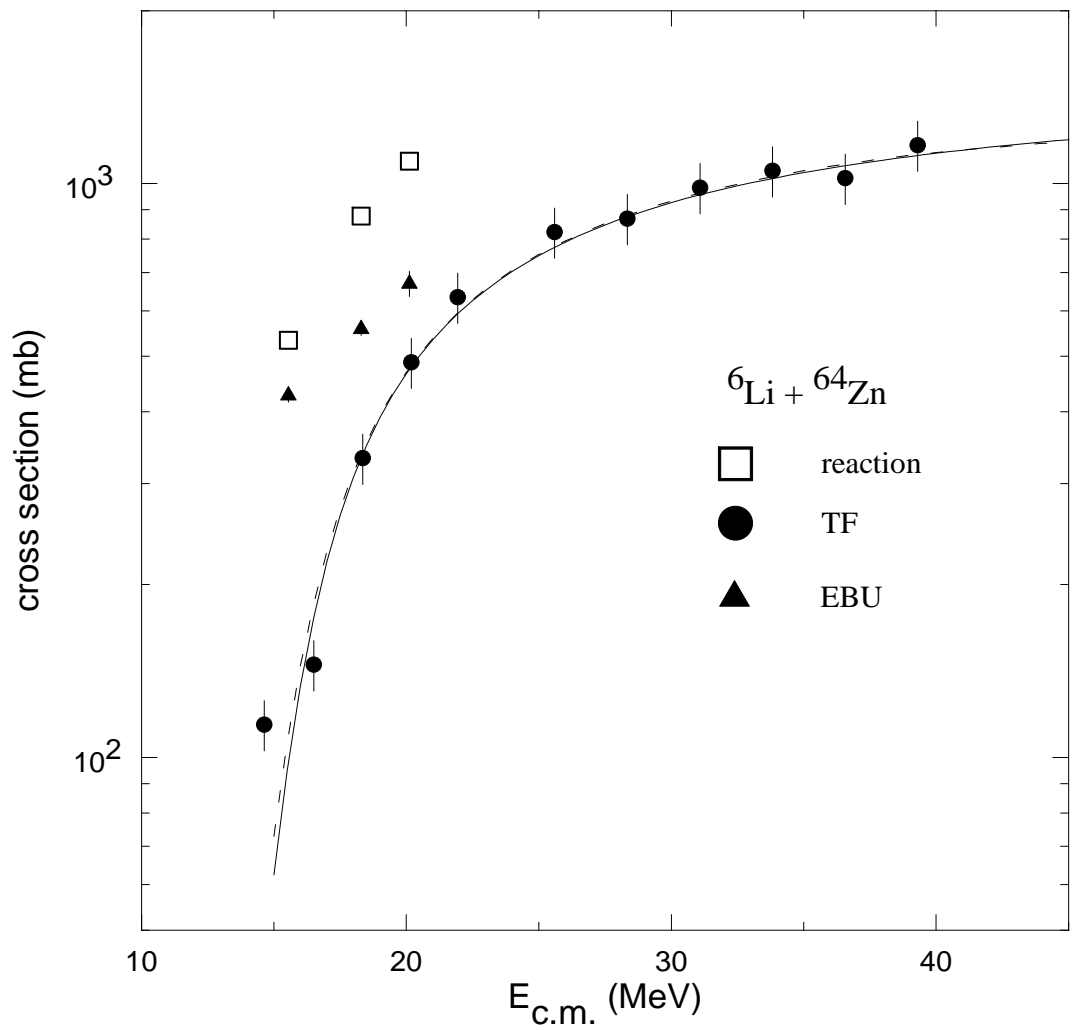
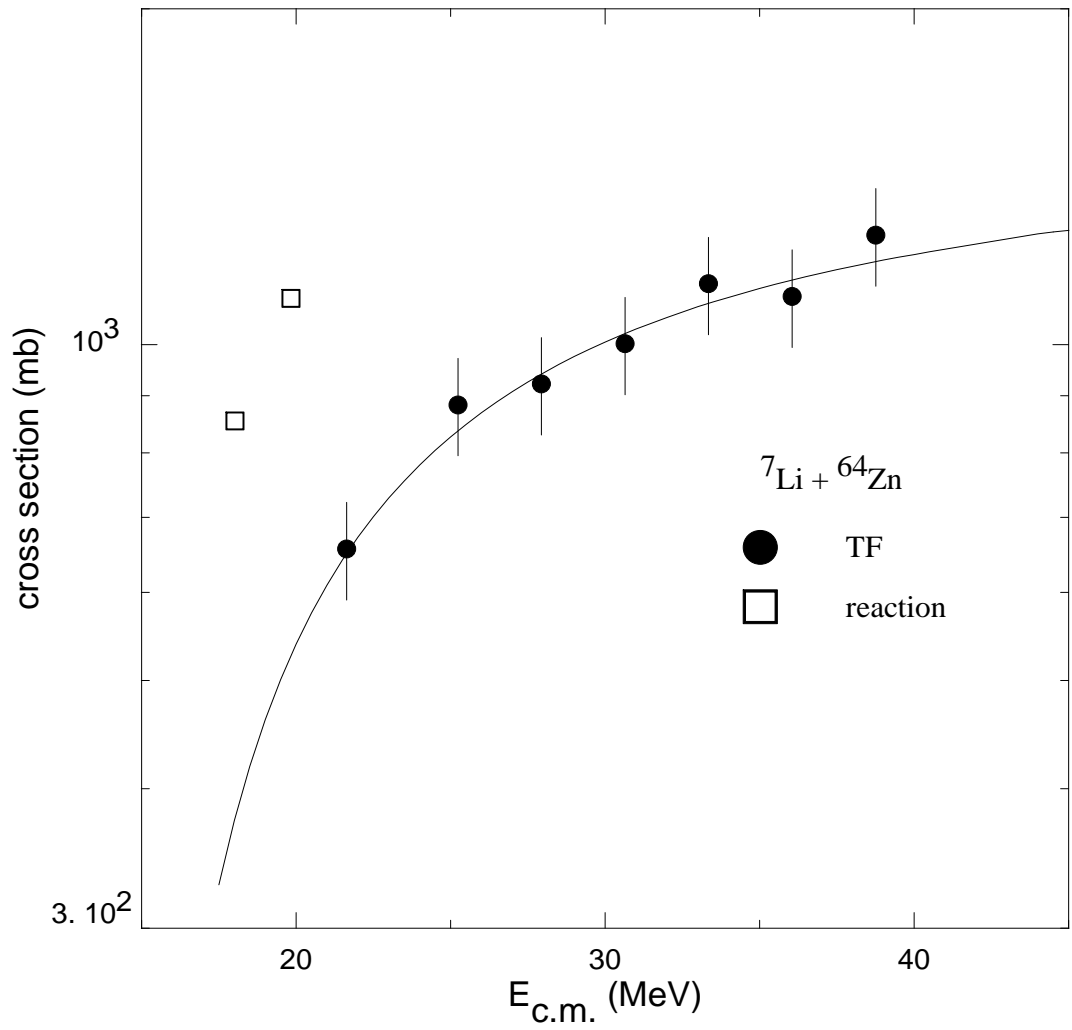
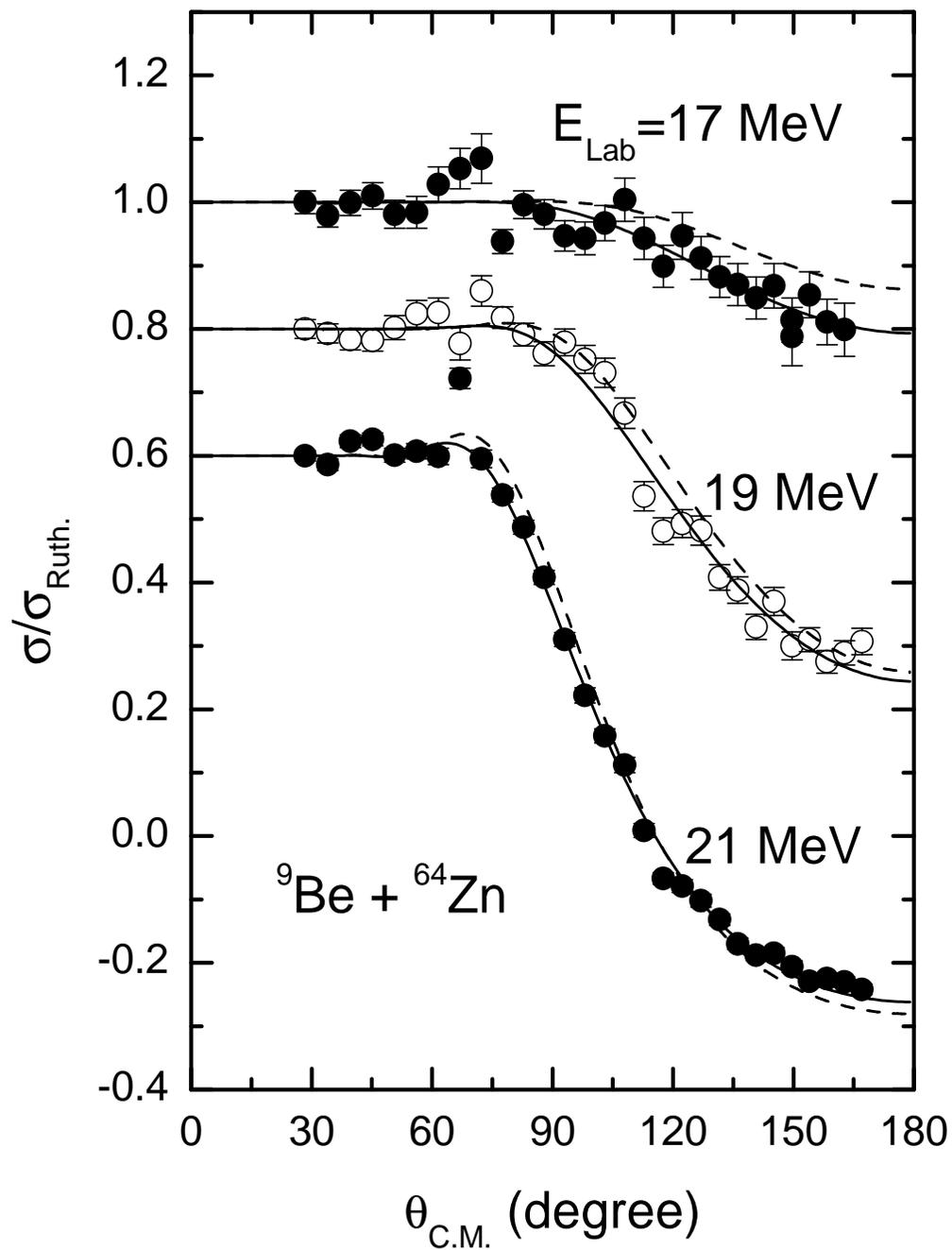
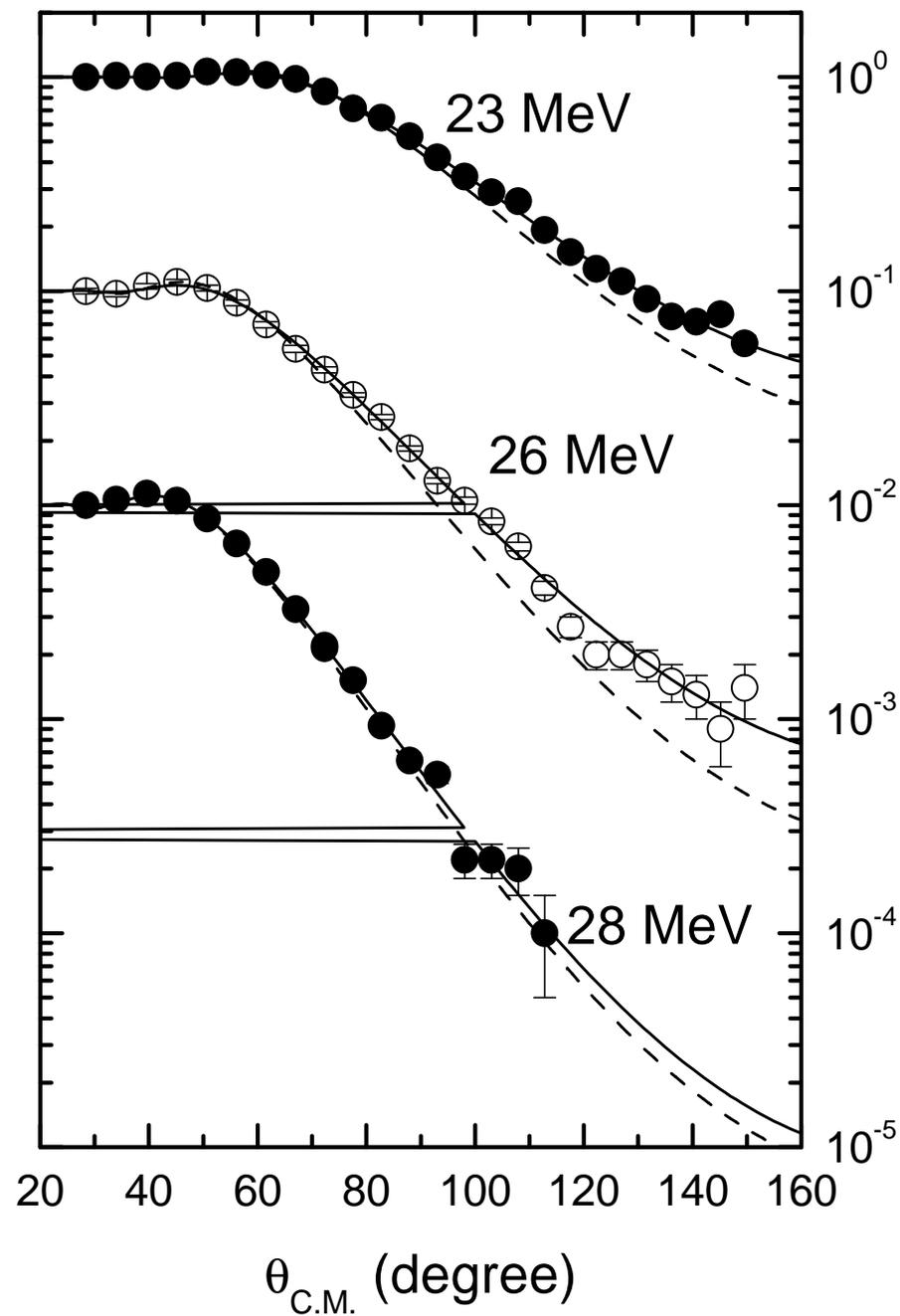
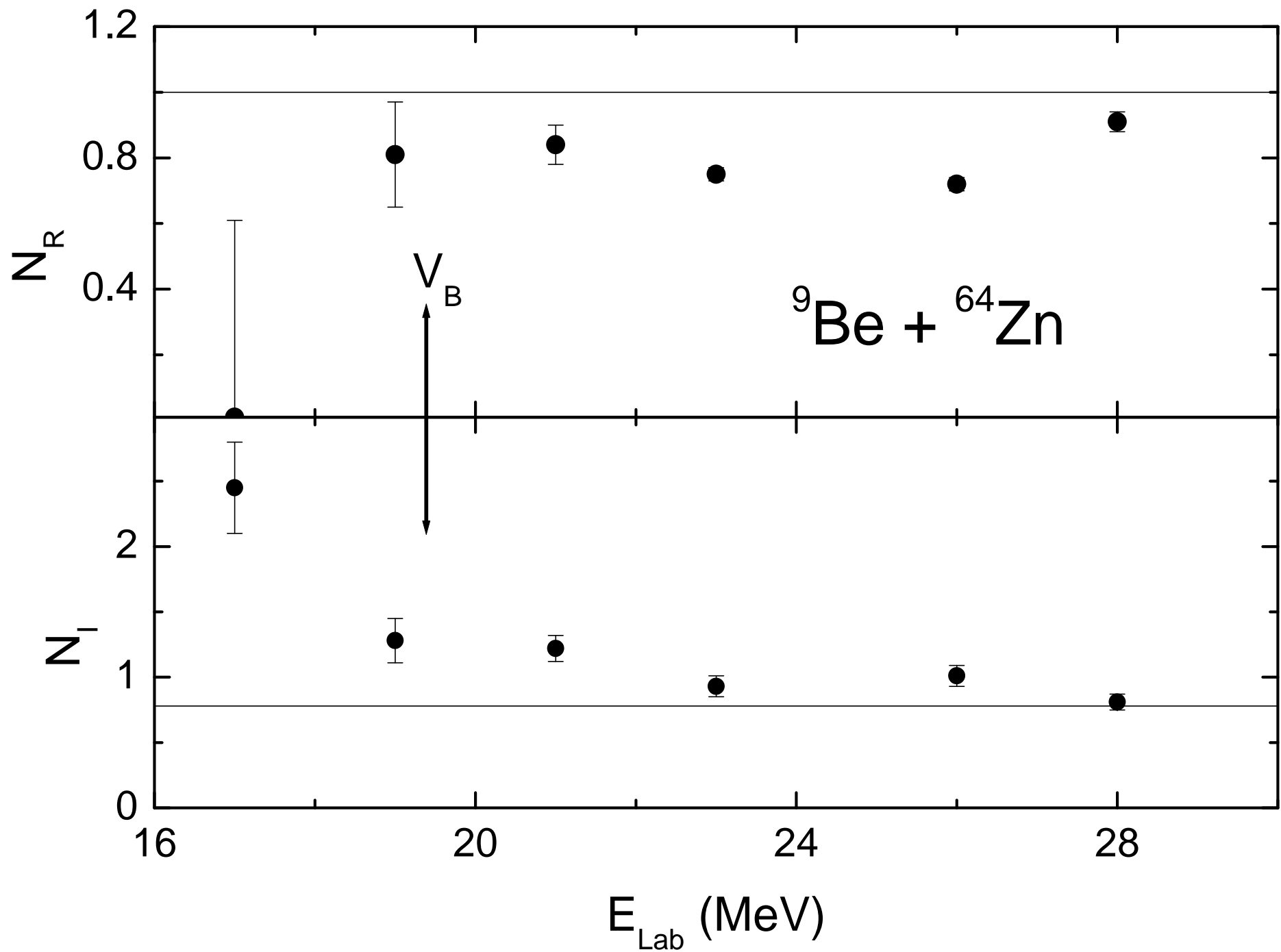
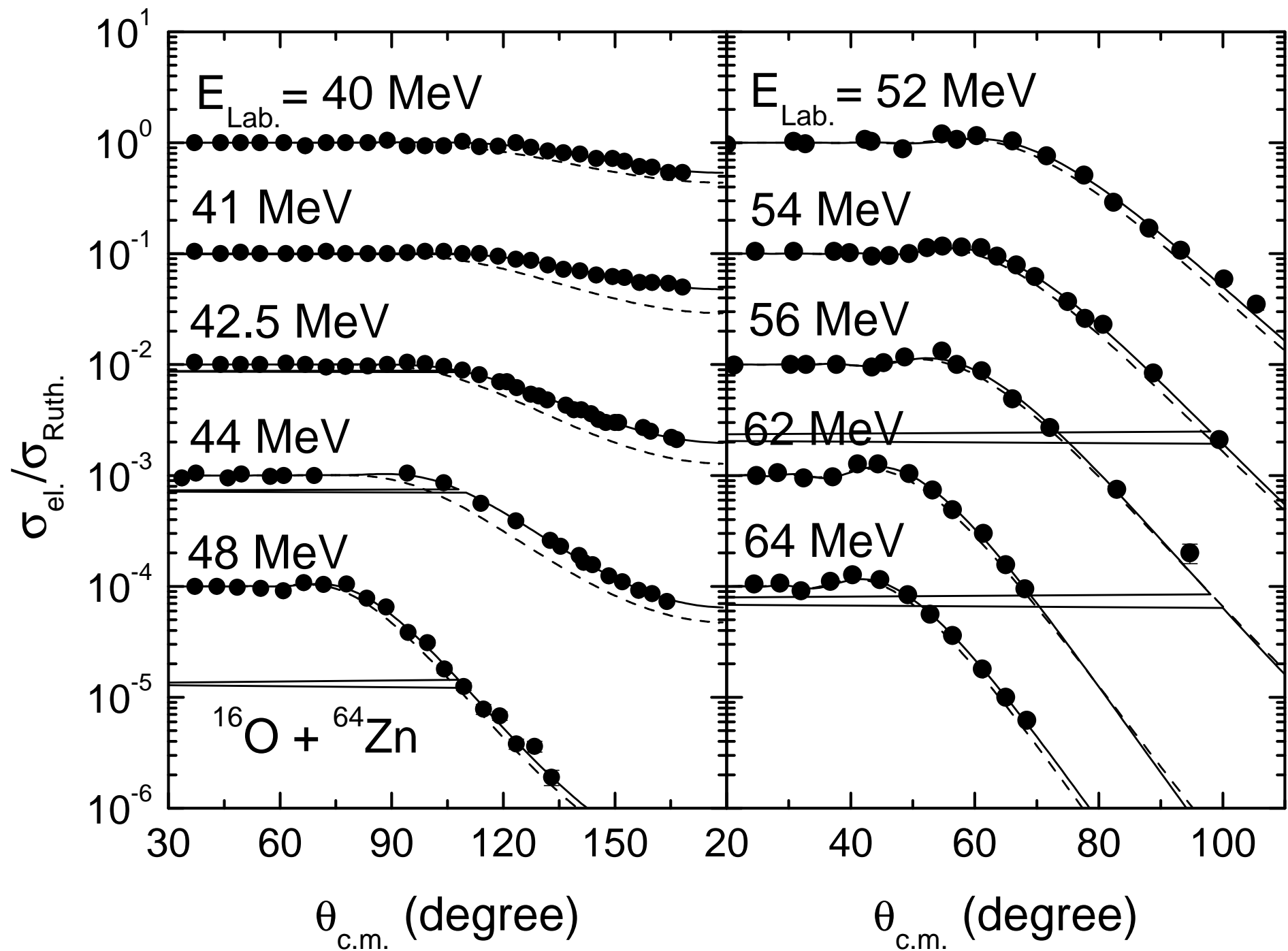
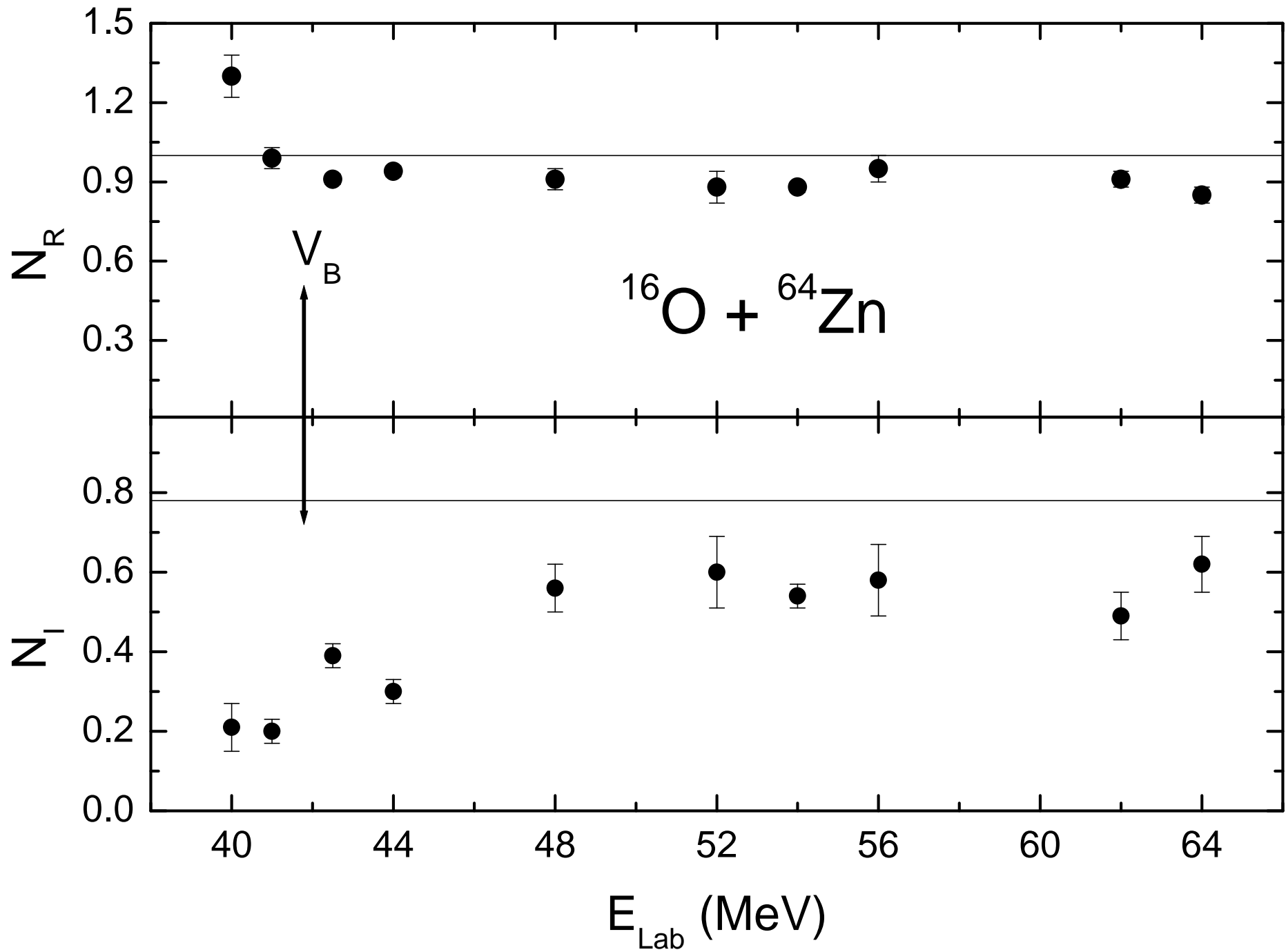
\end{document}